\documentclass[12pt]{iopart}
\usepackage{iopams}  
\usepackage{graphicx}

\begin{document}

\title[]{Reconstructing the final state of Pb+Pb collisions at $\sqrt{s_{NN}}=2.76$~TeV}
\author{Ivan Melo$^{1,2}$ and Boris Tom\'a\v{s}ik$^{2,3}$}
\address{%
$^1$ \v{Z}ilinsk\'a Univerzita, Akademick\'a 1, 01026 \v{Z}ilina, Slovakia}
\address{%
$^2$ Univerzita Mateja Bela, Tajovsk\'eho 40, 97401 Bansk\'a Bystrica, Slovakia
}
\address{%
$^3$ Czech Technical University in Prague, FNSPE, B\v{r}ehov\'a 7, 11519 Prague, 
Czech Republic
}

\date{\today}

\begin{abstract}
We fit the single-hadron transverse-momentum spectra measured in Pb+Pb collisions 
at $\sqrt{s_{NN}} = 2.76$~TeV with the blast-wave model that includes production via 
resonance decays. Common fit to pions, kaons, (anti)protons, and lambdas yields centrality 
dependence of the freeze-out temperature and transverse expansion velocity. In the most 
central collisions we see $T=98$~MeV and $\langle v_t \rangle = 0.654$. The $K^*$
resonance  fits into this picture but the $\phi$ meson might freeze-out a little earlier. 
Multistrange baryons seem to decouple at higher temperature and weaker transverse flow. 
Within our model we observe hints of chemical potential for the charged pions.  
\end{abstract}



\section{Introduction}
\label{intro}

Hot nuclear matter created in ultrarelativistic heavy-ion collisions exhibits its 
properties through the collective expansion of the fireball. In addition to the 
longitudinal flow which is largely set by the initial transparency of the nuclei, pressure
gradients drive strong expansion also perpendicularly to the collision axis. 
Due to expansion and cooling the fireball eventually arrives to the state where it
disintegrates. This is the moment when most abundant hadrons obtain their observed 
momenta and momentum spectra are frozen-out. In fact, it does not happen 
instantaneously for all hadrons. Freeze-out is a continuous process. 

It is useful to describe the freeze-out state of the fireball by hydrodynamically 
inspired parametrisations. They usually assume that freeze-out happens 
instantaneously along some hypersurface. The models include locally thermalized
momentum distribution and some pattern of longitudinal and transverse expansion 
flow velocity. A few parametrisations are on the market 
\cite{Siemens:1978pb,Lee:1988rd,Schnedermann:1993ws,Csorgo:1995bi,%
Tomasik:1999cq,Retiere:2003kf} which differ in details of the 
assumptions for the density and velocity profiles or the exact shape of the freeze-out 
hypersurface. In our analysis we shall make use of the so-called blast-wave model 
\cite{Schnedermann:1993ws}. 

An important part of final state hadrons does not originate directly from the fireball  
but comes from decays of unstable resonances which themselves were emitted from 
the fireball. Hadron production by resonance decays is clearly substantial for pions. 
However,  it also contributes largely to protons and kaons, as we will show below. 
In spite of this, 
it is often omitted in calculations and analyses due to its computational complexity
\cite{Sollfrank:1990qz,Sollfrank:1991xm,Wiedemann:1996ig}.

In this paper we compare Monte-Carlo-generated transverse momentum spectra with 
the data measured in Pb+Pb collisions at $\sqrt{s_{NN}} = 2.76$~TeV by ALICE
collaboration \cite{ALICE_piKp}. Our simulation follows the prescription of the blast-wave 
model. The Monte Carlo treatment has been chosen because in this way we are able to 
include also the resonance production of hadrons into the simulation. 
A similar analysis of RHIC spectra can be found in \cite{k-fit}.

This allows us to deduce the values of kinetic freeze-out temperature and transverse 
expansion velocity much more reliably than in fits with only direct thermal 
production included, e.g.~\cite{ALICE_piKp}.

In the next Section we briefly summarize the relevant features of our Monte Carlo 
model---the package DRAGON \cite{DRAGON}. 
In Section \ref{s:data} we study how  resonance production of 
hadrons affects $p_t$ spectra. As a cross-check of our approach 
we have also performed an analysis  identical to \cite{ALICE_piKp} and then show how results change if 
resonances are included. Our main results are presented in Section \ref{s:results}. 
We have fitted $p_t$ spectra of pions, kaons, (anti)protons, and $\Lambda$'s 
\cite{ALICE_Lambda}
and also looked at $\Xi$'s and $\Omega$'s \cite{ALICE_multistrange}. 
From the resulting fireball freeze-out parameters we calculated spectra of $K^*$ and $\phi$ and 
compared them with the measured data \cite{Abelev:2014uua}.  
We show our values of the parameters
for six different centrality classes. Multistrange baryons seem to drop out of the 
systematics that describes other hadrons. Conclusions are 
drawn in Section~\ref{s:conc}.


\section{The model}
\label{s:model}

Our analysis is performed in framework of the blast-wave model, which is 
characterised by its emission function. Formally, it is the Wigner phase-space 
density of the source of hadrons of type $i$
\begin{eqnarray}
\nonumber
S(x,p)\, d^4x & = & g_i \,  \frac{m_t\, \cosh(\eta-y)}{(2\pi)^3}  \left ( \exp\left ( \frac{p_\mu u^\mu - \mu_i }{T} \right ) + s_i \right )^{-1}
\theta\left ( 1 - \frac{r}{R} \right )
\nonumber \\
& & \qquad \qquad
\times r\, dr\, d\varphi\, \delta(\tau - \tau_0)\, 
\tau\, d\tau\, d\eta\,  . 
\label{e:Sfun}
\end{eqnarray}
Because of the dominant expansion in the longitudinal direction one uses longitudinal 
proper time $\tau = \sqrt{t^2 - z^2}$ and space-time rapidity 
$\eta = \frac{1}{2}\ln((t+z)/(t-z))$. Polar coordinates $r$, $\varphi$ are used in the 
transverse plane. We use proper quantum statistical distributions with $s_i = 1$ 
$(-1)$ for fermions (bosons) and $g_i$ is the spin degeneracy. Every isospin state 
is treated separately. This prescription assumes sharp freeze-out along the hypersurface
$\tau = \tau_0$ and uniform density distribution within the radius $R$. 
This means that our freeze-out time does not depend on radial coordinate\footnote{%
This is one of the differences between our model and the Cracow single freeze-out 
model \cite{Broniowski:2001we} which was used in fits to $p_t$ spectra measured by 
ALICE collaboration recently \cite{Begun:2013nga,Begun:2014rsa,Chatterjee:2014lfa}.
}.
The expansion of 
the fireball is represented by the velocity field
\begin{equation}
u^\mu  = \left (  \cosh\eta_t \cosh\eta,\, \sinh\eta_t\cos\varphi, 
 \sinh\eta_t \sin\varphi,\, \cosh\eta_t\sinh\eta\right )
\end{equation}
where the transverse velocity is such that 
\begin{equation}
v_t = \tanh\eta_t = \eta_f \left ( \frac{r}{R} \right )^n\,  .
\end{equation}
In this relation $\eta_f$ parametrises transverse flow gradient and $n$ the profile 
of the transverse velocity. The mean transverse velocity is then
\begin{equation}
\label{e:meanv}
\langle v_t \rangle = \frac{2}{n+2} \eta_f\,  .
\end{equation}
The transverse size $R$ and the freeze-out proper time $\tau_0$ influence total 
normalizations of transverse momentum spectra. However, in this study we ignore 
those and hence we have no sensitivity to these geometric parameters.

From the emission function, spectrum of directly produced hadrons is obtained as
\begin{equation}
\label{e:sspec}
E\frac{d^3N}{dp^3} = \int_\Sigma S(x,p)\, d^4x\,  ,
\end{equation}
where the integration runs over the whole freeze-out hypersurface. If one replaces 
the quantum-statistical distribution in eq.~(\ref{e:Sfun}) by the classical Boltzmann
distribution and performs some of the integrations in eq.~(\ref{e:sspec}), one arrives at 
\begin{equation}
E\frac{d^3N}{dp^3} = \frac{m_t \tau_0}{2\pi^2} e^{\mu_i/T}  
\int_0^R r\, I_0\left ( \frac{p_t}{T}\sinh\eta_t(r) \right ) \,
 K_1\left ( \frac{m_t}{T}\cosh\eta_t(r) \right )\, dr\,  .
\label{e:dirspec}
\end{equation}
This formula is rather easy to evaluate and thus it is often used in spectra fitting. 

Resonances are emitted as described by the emission function in eq.~(\ref{e:Sfun}) and 
then decay exponentially in time according to their width. All matrix elements for 
the decays are assumed to be constant and thus the decay is determined by the 
phase-space only. In principle, one can derive the emission function that includes
hadrons from resonance decays \cite{Schnedermann:1993ws}  
but the expression---unlike eq.~(\ref{e:dirspec})---is 
not suitable for massively multiple use in the fitting procedure. 

Therefore, we generate stable hadrons and also resonances according to distribution 
$S(x,p)$ and the assumption of chemical equilibrium. 
Resonances are then let to decay so that in the end we look only at stable hadrons. 
In our simulation we will 
systematically vary the temperature $T$, transverse flow gradient $\eta_f$, 
and the power $n$. We construct histograms in $p_t$ for 
some number of Monte Carlo events for each set of $(T,\eta_f,n)$ parameters 
which are then compared with measured data and a value of $\chi^2$ is obtained. 
This procedure is rather CPU and storage-demanding because the generation must be done separately for 
each set of parameters that we want to check. 

The share of hadrons coming from resonance decays in our model is set at the chemical  
freeze-out specified by $T_{ch} = 152$~MeV and $\mu_B=0$ 
\cite{Milano:2013sza}. Note that shifting these values slightly is not expected to cause
a big change in the \emph{shape} of the transverse momentum spectra.

This value of chemical freeze-out temperature was found to reproduce the \emph{ratios}
of hadron multiplicities \cite{Milano:2013sza}
and we keep it in order to satisfy that observation. 
Since the thermal freeze-out temperature is lower, we actually assume here 
that the time between chemical and thermal freeze-out is so short that the majority 
of resonances does not decay so that their products would rescatter. The opposite extreme 
scenario (not studied here) is that strong interactions regenerate some number of resonances. 
This must happen so that the state of \emph{partial chemical equilibrium} is established,
because the final observed numbers of stable hadrons is fixed to the values that 
correspond to the chemical freeze-out state. Thus in partial chemical equilibrium the loss 
of pions due to fewer resonances that decay would have to be exactly compensated by the 
increase of the number of direct pions. The real scenario may likely be somewhere 
in between these two extreme models. 
Another possibility is the scenario with 
single chemical and thermal freeze-out \cite{Begun:2013nga,Begun:2014rsa}. 
As mentioned above, due to fast transverse expansion at the LHC, here we assume only short
time between chemical and thermal freeze-out which keeps basically all resonances 
according to the chemical freeze-out temperature. 
At the chemical freeze-out, all hadrons are produced from the same common volume. 
Then, in the short time of further expansion the volume occupied by each species may be
slightly different due to the difference in their thermal velocities.
Same kind of model 
has been applied also in an analysis of nuclear collisions at RHIC energies \cite{k-fit}. 


\section{Impact of resonance decays on transverse momentum spectra}
\label{s:data}

Data from Pb+Pb collisions at $\sqrt{s_{NN}} = 2.76$~TeV were measured 
by the ALICE collaboration. Spectra of the most abundant species---pions, kaons, 
protons and antiprotons---were published in \cite{ALICE_piKp} along with a simple thermal 
fit to the spectra which included only directly produced particles and no resonance decay effect. 
An effort has been made to identify fiducial intervals in $p_t$ where the influence 
of resonance decays can be omitted. They were  $0.5~\mathrm{GeV}/c<p_t<1~\mathrm{GeV}/c$ for 
pions, $0.2~\mathrm{GeV}/c<p_t<1.5~\mathrm{GeV}/c$ for kaons and $0.3~\mathrm{GeV}/c<p_t<3~\mathrm{GeV}/c$ 
for protons. As we first wanted to set the benchmark and compare our method to other analyses, 
we performed the same blast-wave model fit with the same fiducial intervals as \cite{ALICE_piKp} 
using our Monte Carlo approach instead of eq.~(\ref{e:dirspec}). 

Event samples were generated for different sets of
parameters $(T,\eta_f,n)$. The parameter space was sampled so that the step in temperature 
was 4~MeV, the step in $\eta_f$ was 0.01, and the step in $n$ was 0.02 or 0.04. Each of these steps 
induces roughly the same amount of change in the resulting spectra. The step size was compromised 
in view of the CPU time required and has impact on the precision with which we can locate 
the parameter set that fits the data best. Explored parameter sets were chosen so that 
we always explore large enough region in parameter space around the best fit. 

We checked that the generated spectra are accurately described by eq.~(\ref{e:dirspec}) with appropriate parameters. 
Then we looked for the parameter set $(T,\eta_f,n)$ for which Monte Carlo data show the best match with 
measured data using the minimum $\chi^2$ method. Results are summarised in Table~\ref{t:alicefits}.
They agree with those by ALICE collaboration \cite{ALICE_piKp} reasonably well.

\begin{table}
\caption{Freeze-out parameters obtained from fits to pion, kaon and proton $p_t$ spectra
within fiducial intervals as defined in \cite{ALICE_piKp}. Results without and with 
resonance decays included are compared to the results by ALICE collaboration. 
\label{t:alicefits}}
\begin{indented}
\lineup
\item[]\begin{tabular}{llllllllll}
\br
 & \centre{3}{ALICE \cite{ALICE_piKp}} & \centre{3}{no resonances} & 
\centre{3}{with resonances} \\ \ns\ns
& \crule{3} & \crule{3} & \crule{3}\\
centrality & \parbox{2.7em}{$T$ (MeV)} & $\langle v_t \rangle$ & $n$ &
\parbox{2.7em}{$T$ (MeV)} & $\langle v_t \rangle$ & $n$ & 
\parbox{2.7em}{$T$ (MeV)} & $\langle v_t \rangle$ & $n$\\
\mr
\00--5\% & \095 & 0.651& 0.71& \098& 0.645& 0.73& \082& 0.662& 0.69\\
\05--10\% & \097 & 0.646 & 0.72 & \098 & 0.645 & 0.73 & \094 & 0.654 & 0.69\\
10--20\% & \099 & 0.639 & 0.74 & 102 & 0.637 & 0.73 & \090 & 0.649 & 0.71 \\
20--30\% & 101 & 0.625 & 0.78 & 102 & 0.624 & 0.79 & \098 & 0.633 & 0.75 \\
30--40\% & 106 & 0.604 & 0.84 & 110 & 0.605 & 0.81 & 102 & 0.616 & 0.79 \\
40--50\% & 112 & 0.574 & 0.94 & 110 & 0.572 & 0.97 & 118 & 0.581 & 0.89 \\
50--60\% & 118 & 0.535 & 1.10 & 122 & 0.527 & 1.15 & 126 & 0.541 & 1.03\\
60--70\% & 129 & 0.489 & 1.29 & 126 & 0.484 & 1.39 & 146 & 0.489 & 1.23 \\
70--80\% & 139 & 0.438 & 1.58 & 142 & 0.439 & 1.51 & 170 & 0.423 & 1.55 \\
\br
\end{tabular}
\end{indented}
\end{table}

Then, in order to check whether resonances can indeed be neglected in the 
fiducial range, we added resonance decays to our Monte Carlo data. 
Their contribution is characterised by the \emph{chemical} freeze-out temperature
as described in the previous section. 
We still kept 
the limitation to the fiducial $p_t$ intervals. 
Results are summarised in Table~\ref{t:alicefits}.
The extracted kinetic freeze-out temperatures changed  and the direction in which 
they have gone depends on centrality. For central collisions the temperature 
was lowered while for peripheral ones it went up. 
It thus seems that leaving out resonances is not a good approximation even if 
one tries to identify a reasonable fiducial range for the fitting. 

To support this statement we have analysed our Monte Carlo events and looked at 
the number of hadrons emitted directly and the number coming from the decays of resonances. 
\begin{figure}[t]
\begin{indented}
\item[]\includegraphics[width=0.8\textwidth]{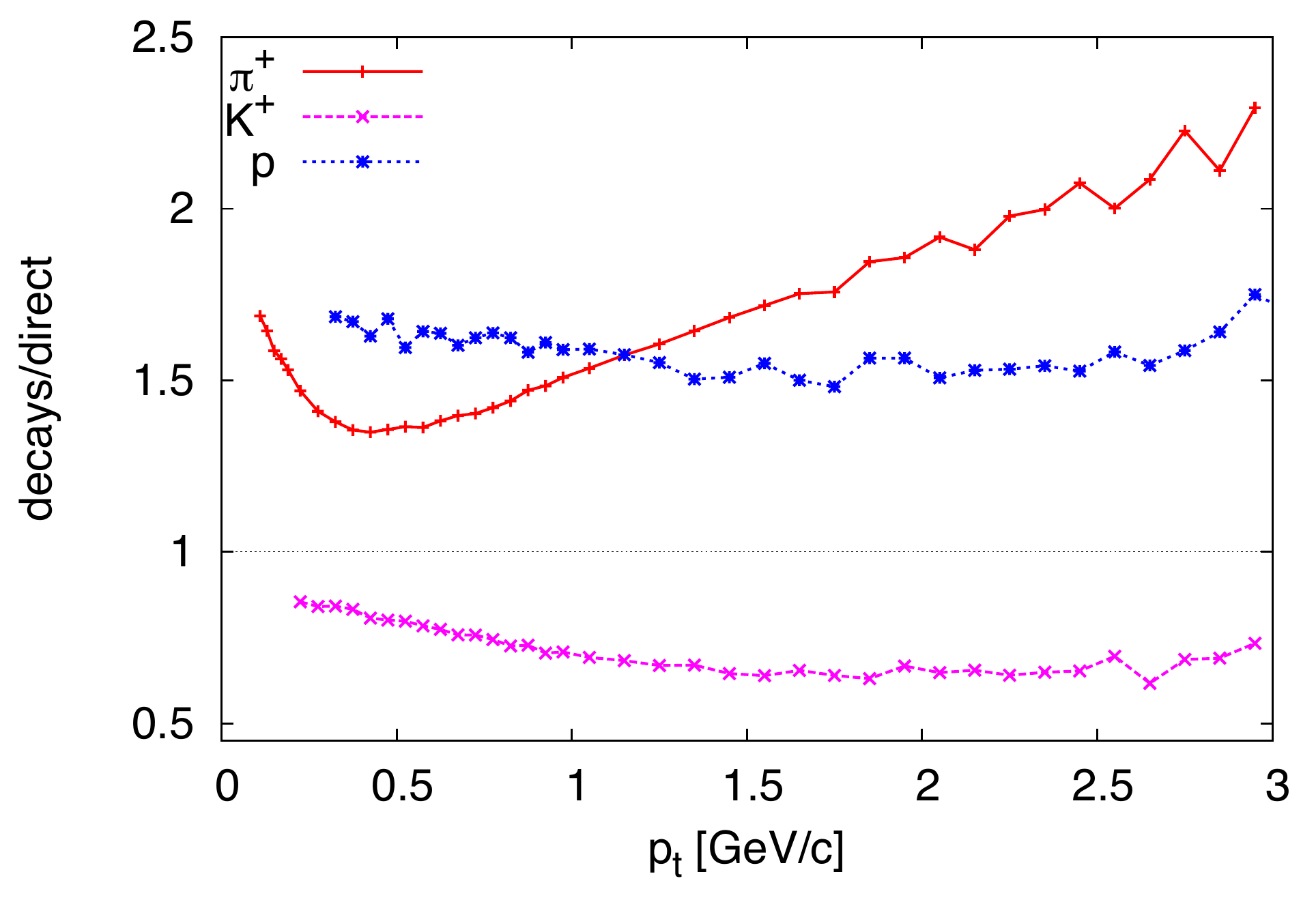}
\end{indented}
\caption{\label{f:dirres}%
Ratios of hadrons coming from resonance decays to those produced
directly for three different species. Monte Carlo data were simulated by DRAGON 
with the parameters $T = 98$~MeV, $\eta_f = 0.88$, and $n = 0.69$.
}
\end{figure}

In Fig.~\ref{f:dirres} we plot the ratio of hadrons produced from the resonance decays to 
those produced directly as functions of $p_t$. They are calculated for model 
parameters which best fit the spectra in most central collisions (see Table~\ref{t:bestfits}).
The portion of resonance production is very 
important not only for pions but also for kaons and protons. Moreover, the ratio shows 
nontrivial $p_t$ dependence. One might omit resonances in the 
fitting procedure in a region where these ratios are constant but there is no 
such interval for any of the investigated species. For pions the $p_t$-dependence
is even non-monotonic. Hence, leaving out resonances is an invalid simplification. 

\begin{figure}[t]
\begin{indented}
\item[]\includegraphics[width=0.8\textwidth]{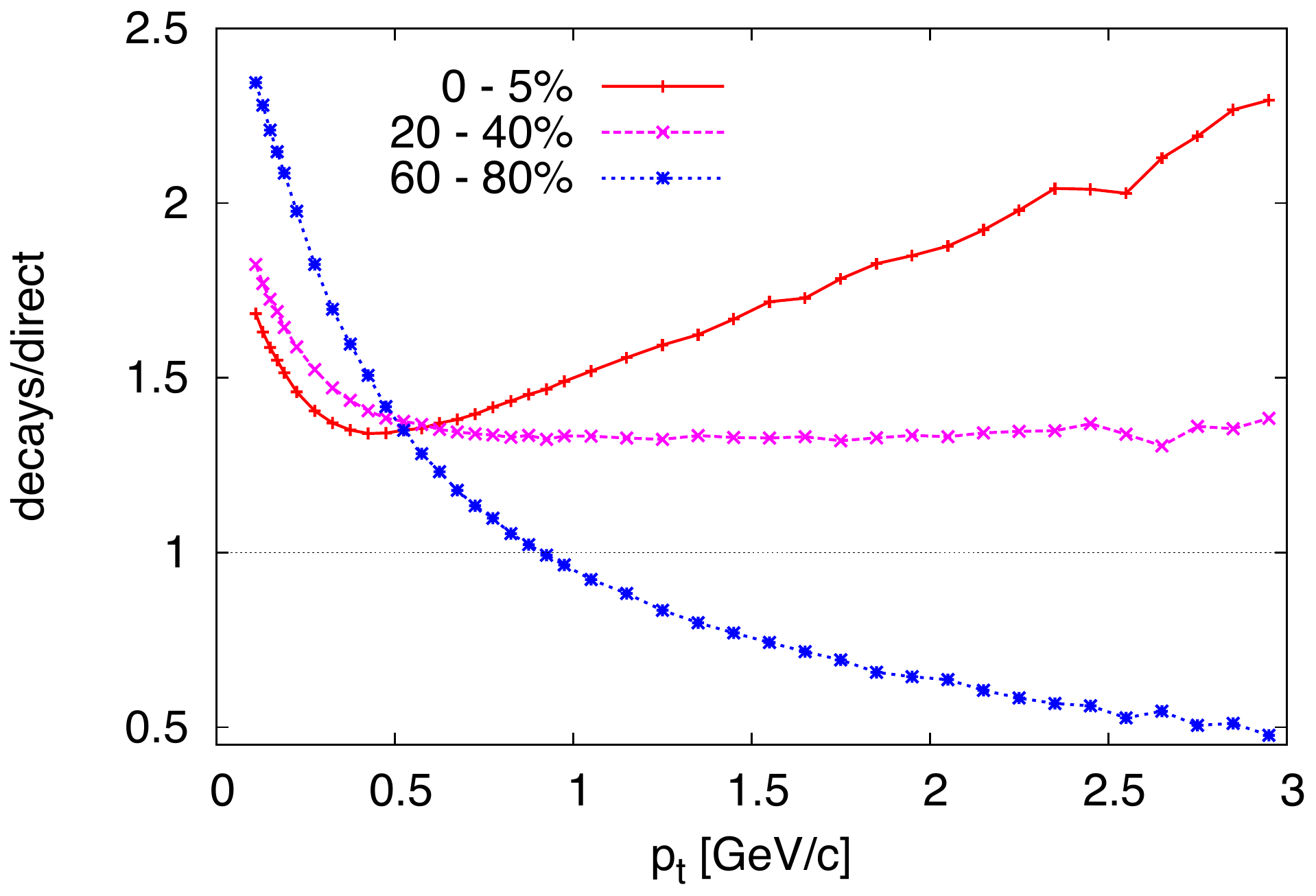}
\end{indented}
\caption{\label{f:pires}%
Ratios of pions coming from resonance decays to those produced
directly. Monte Carlo data were simulated by DRAGON 
with the parameters given by the best fits in the indicated centrality class. 
}
\end{figure}
We analysed the resonance contribution to hadron spectra in much detail. In 
Fig.~\ref{f:pires} we demonstrate that the relative contribution of resonance decays to 
pion production also depends on kinetic freeze-out temperature and transverse flow. 
Simulated data correspond to best fits in the indicated centrality classes (next Section). 
The rise of decays/direct ratio with increasing $p_t$ is due to decays of resonances 
with masses so high that pions get large kinetic energy from their decays. In central 
collisions the freeze-out temperature is low. Thus at higher $p_t$ the pion spectrum 
is dominated by heavy resonance decay products. It also helps  
that these resonances are boosted due to large transverse velocity of the part of fireball 
which emits them. In peripheral collisions the freeze-out temperature increases and 
the transverse flow is weaker. The former gives high $p_t$ to direct pions and the 
latter gives less boost to heavy resonances. As a result, the contribution of resonance 
production decreases with growing $p_t$. 


\section{Fits to hadron transverse momentum spectra}
\label{s:results}

In our actual analysis we fitted spectra of pions, charged kaons, 
protons and antiprotons \cite{ALICE_piKp} together with spectra of $K^0$'s and $\Lambda$'s \cite{ALICE_Lambda}. 
Our simulations were further compared with spectra of multistrange baryons $\Xi$ and $\Omega$ published in 
\cite{ALICE_multistrange}. Finally, we compared with data \cite{Abelev:2014uua} our predictions
for $K^*$ and $\phi$ which were based on the results of fits to $\pi$, $K$, $p$, $\bar p$, and $\Lambda$. 

The $K^0$ and hyperon data are divided into fewer centrality classes than data from 
$\pi^\pm$, $K^{\pm}$, 
$p$, and $\bar p$. In order to make consistent fits which include all species, we used the 
centrality classes of hyperon spectra and summed up corresponding centrality 
bins in pion, kaon, and proton spectra. Error bars of experimental data points were taken as
statistical and systematic added in quadrature.

The parameter space $(T,\eta_f,n)$ was sampled as described in the previous section.
After having all ensembles of events simulated we compared them with data.  Histograms 
of simulated data were normalized individually so that their normalization matches that 
of experimental spectrum. Among all simulated ensembles we found the one which 
best agrees with the data (the minimum $\chi^2$). 
We chose this procedure in order to find the best fits to the \emph{shape} of the spectrum 
which is sensitive to the dynamical state of the fireball at the freeze-out. Their normalization, 
on the other hand, requires the volume which is best reconstructed in common analysis with 
femtoscopy data. 


Fit results depend on the cuts imposed on the spectra.
For each species we specified a high-$p_t$ cut. This is justified, because from some 
value of $p_t$ upwards hadrons are expected to be more influenced by hard production 
processes and not described by the thermal model we use. 
To find the cuts we determined ratios of measured-to-simulated bin content
$R_i = N_i^{\mathrm{exp}}/N_i^{\mathrm{MC}}$ for each species $i$ as a function of
$p_t$. Then, those bins from the high-$p_t$ end of the spectra for which the value of $R_i$ was outside the range (0.9, 1.1) 
were excluded, effectively introducing high-$p_t$ cuts. (The chosen range was found to lead to best fits.)
In the next step, Monte Carlo data were again compared with the data without excluded bins. 
This time, new high-$p_t$ cuts were determined and the procedure 
was repeated again. The iterations were stopped when they converged to one set of parameters
and certain set of cuts.

For pions, we imposed also a low $p_t$ cut at 400~MeV/$c$, thus 
not including the first 9 bins into the comparison. We found that the quality of the fit significantly deteriorated without this cut.
It seems that pion spectra show an enhancement here which might be due to non-equilibrium 
pion chemical potential not included in our simulation.

Here we did not try to come up with a fit function using Tsallis distribution that would 
cover the whole $p_t$ interval \cite{Wilk:2008ue,Cleymans:2008mt,Wei:2015oha}. 
Its interpretation in terms of thermal model 
would be unclear. 


\subsection{Spectra of $p,{\bar p}, \pi^{\pm}, K^{\pm}, K^0, \Lambda$}

\begin{figure}[t]
\begin{indented}
\item[]\includegraphics[width=0.85\textwidth]{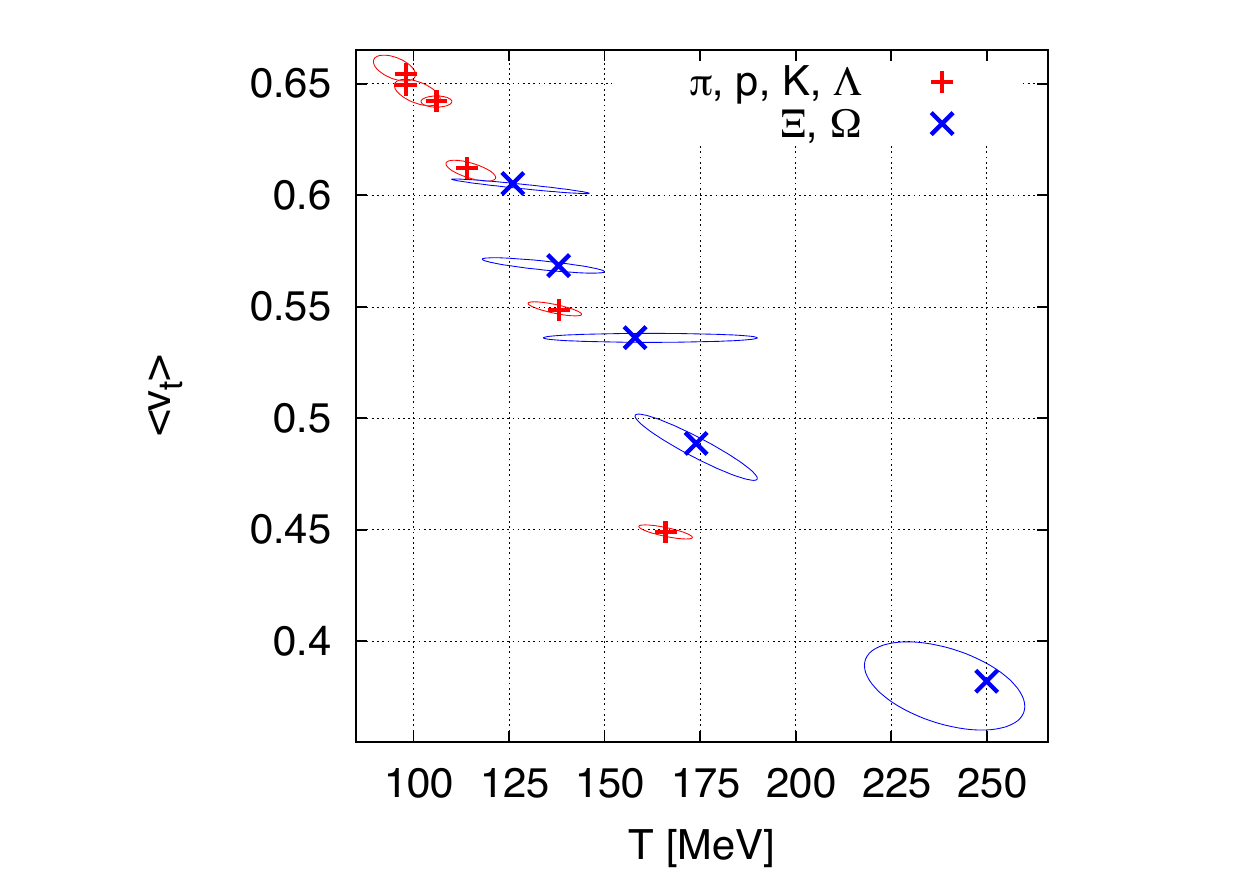}
\end{indented}
\caption{\label{f:bestfit}%
Positions of best-fit values of $T$ and $\langle v_t\rangle$ for 
transverse momentum spectra of $\pi^\pm$, $K^\pm$, $p$, $\bar p$,
$K^0$, and $\Lambda$ are shown by red +'s. Results go from central 
(upper left) to non-central (lower right) collisions, with centrality classes 
indicated in Table~\ref{t:bestfits}. Around each best fit value
we estimate the 99\% confidence-level region. By blue $\times$'s we
show results of fits to spectra of multistrange species, centralities
are indicated in Table~\ref{t:strangeparams}. For multistrange 
species we estimate the 68\% confidence-level regions. 
}
\end{figure}

We first fitted $p_t$ spectra for the most abundant species: pions, kaons, (anti)protons,  
$K^0$'s, and $\Lambda$'s.  
In Fig.~\ref{f:bestfit} we show positions of the best fits for each centrality.
Note that in the actual fitting procedure we determined three parameters: 
$T$, $\eta_f$, and $n$. In order to compare results in a relevant way
we have plotted here the positions of the best fits in just two parameters, $T$ and $\langle v_t \rangle$, 
with the latter calculated from eq.~(\ref{e:meanv}) for the 
values of $n, \eta_f$ that give the minimum of $\chi^2(T,\eta_f,n)$. 
As can also be seen in 
Table~\ref{t:bestfits}, with the change of centrality, 
$\langle v_t \rangle$ sometimes  changes due to 
a change in $n$ and not in $\eta_f$.  
\begin{table}
\caption{The best parameter values resulting from fits to transverse momentum 
spectra of pions, kaons, (anti)protons, $K^0$'s and $\Lambda$'s. 
\label{t:bestfits}}
\lineup
\begin{indented}
\item[]\begin{tabular}{lllllll}
\br
centrality & \parbox{2.7em}{$T$ (MeV)}  & $\eta_f$ & $n$ & $\langle v_t \rangle$ & $\chi^2/N_\mathrm{dof}$ & $N_\mathrm{dof}$ \\
\mr
\00--5\%   & \098 & 0.88 & 0.69 & 0.654 & 0.214 & 194 \\
\05--10\% & \098 & 0.88 & 0.71 & 0.649 & 0.266 & 197 \\
10--20\% & 106 & 0.87 & 0.71 & 0.642 & 0.272 & 210 \\
20--40\% & 114 & 0.86 & 0.81 & 0.612 & 0.294 & 202 \\
40--60\% & 138 & 0.82 & 0.99 & 0.548 & 0.347 & 195 \\
60--80\% & 166 & 0.77 & 1.43 & 0.449 & 0.449 & 168 \\
\br
\end{tabular}
\end{indented}
\end{table}
In Table~\ref{t:bestfits} the fitted values are displayed together with their
$\chi^2$ and the number of degrees of freedom. In Fig.~\ref{f:bestfit}
they are displayed together with our estimates for the uncertainty region for the 
best fit values. We stress that these are only estimates because the evaluation of
$\chi^2$ in discrete points does not allow us to determine them more precisely. 
We observe that the position of minimum determines very precisely the value 
of $\langle v_t \rangle$. For the temperature, the uncertainty is larger. It may range up to 10--15~MeV
from the minimum.

\begin{figure}[t]
\begin{indented}
\item[]\includegraphics[width=0.8\textwidth]{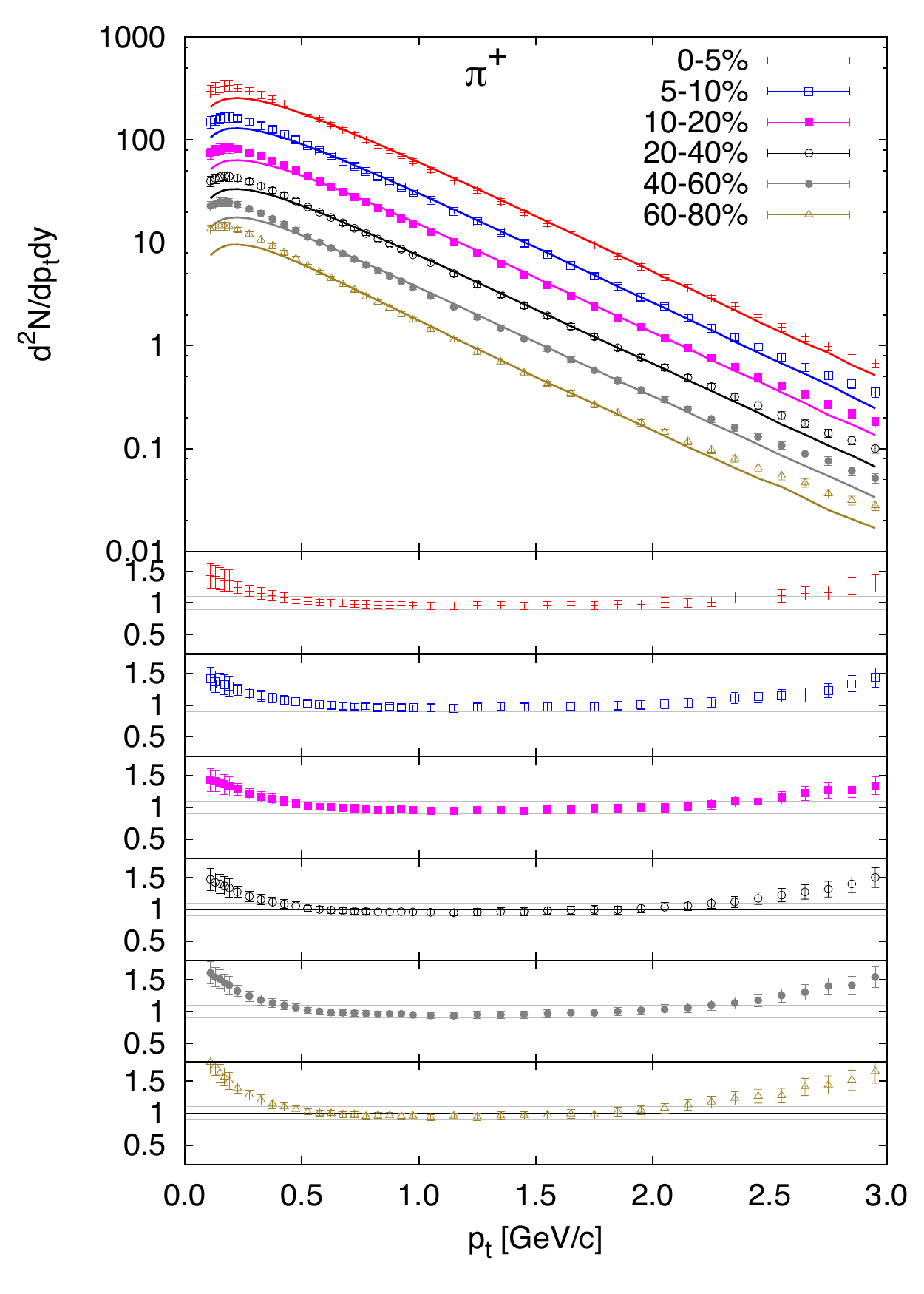}
\end{indented}
\caption{\label{f:pifit}%
Transverse momentum spectra of $\pi^+$ for different centralities. 
To display all spectra in one figure we divide data from non-central collisions
by factors 2, 4, 8,  16, and 32.
Lower panels show the experiment-to-Monte-Carlo ratios 
$R_i = N_i^\mathrm{exp}/N_i^\mathrm{MC}$ which demonstrate the 
agreement of measured and fitted spectra. Different panels correspond to 
different centralities. Horizontal lines indicate the interval 0.9 $\le R_i \le$ 1.1 to 
which we limit our fitting procedure.
}
\end{figure}
\begin{figure}
\begin{indented}
\item[]\includegraphics[width=0.8\textwidth]{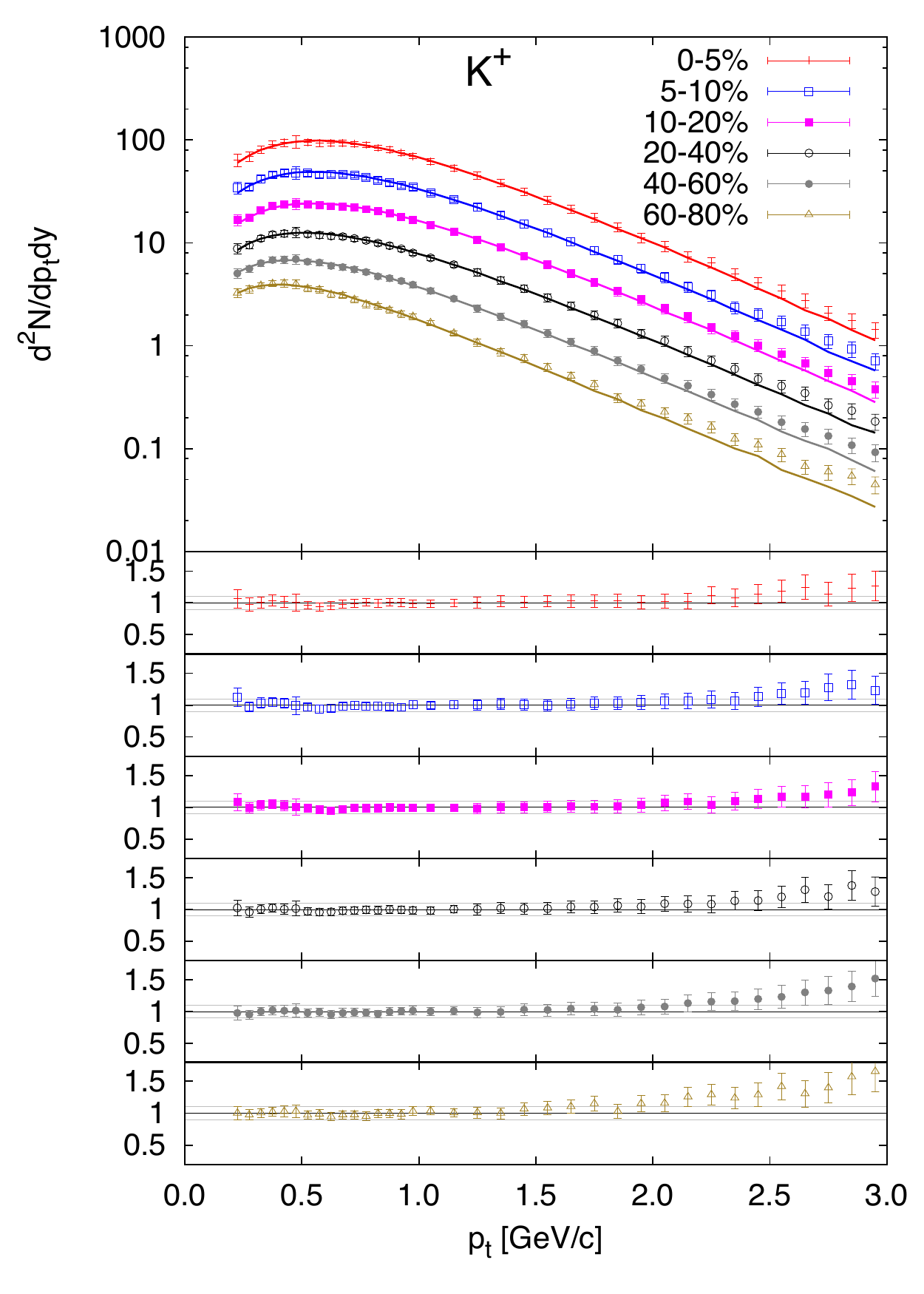}
\end{indented}
\caption{\label{f:Kfit}%
Same as Fig.~\ref{f:pifit}, but for positive kaons. 
}
\end{figure}
\begin{figure}
\begin{indented}
\item[]
\includegraphics[width=0.8\textwidth]{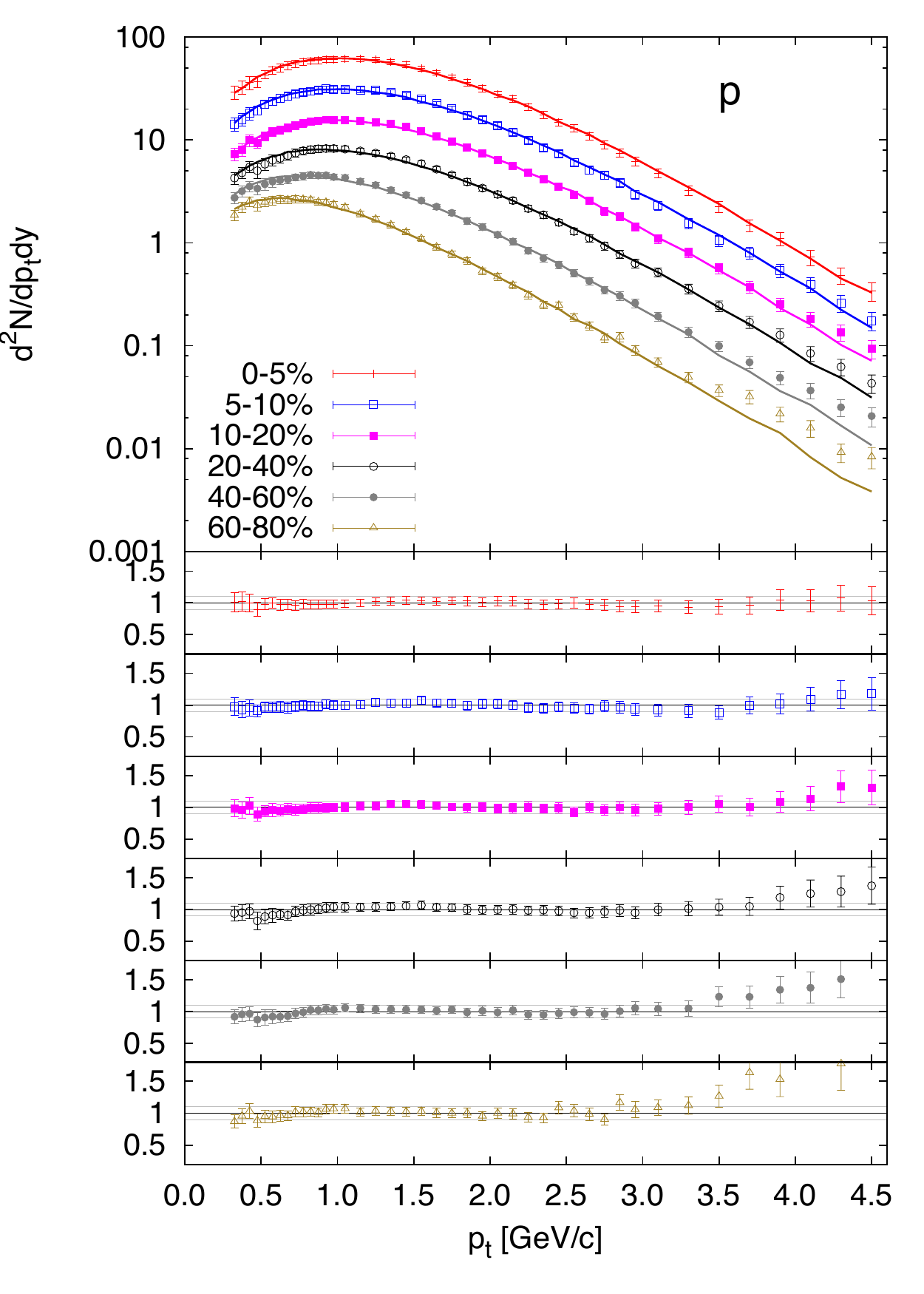}
\end{indented}
\caption{\label{f:pfit}%
Same as Fig.~\ref{f:pifit}, but for protons. 
}
\end{figure}
\begin{figure}
\begin{indented}
\item[]\includegraphics[width=0.8\textwidth]{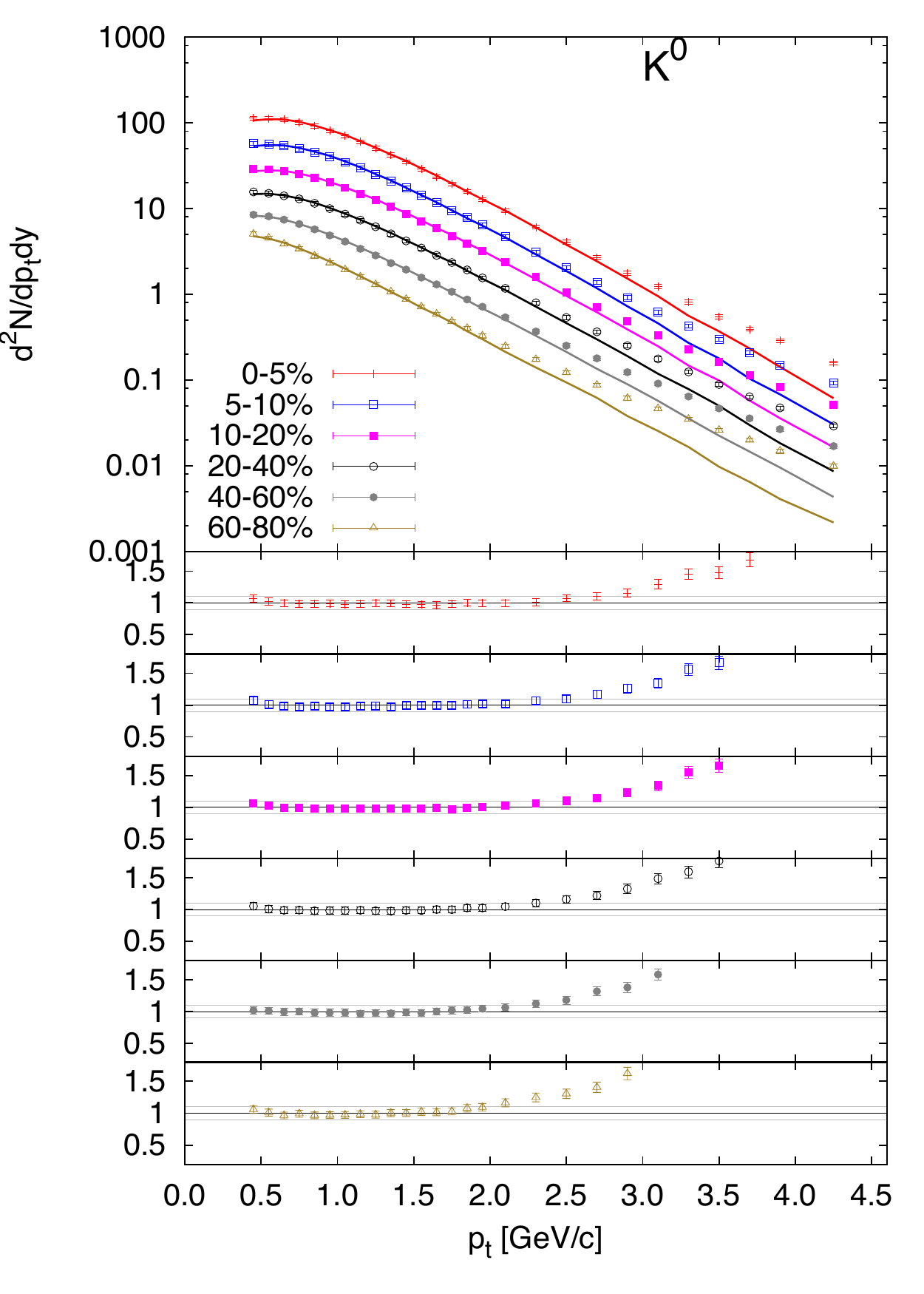}
\end{indented}
\caption{\label{f:K0fit}%
Same as Fig.~\ref{f:pifit}, but for $K^0$. 
}
\end{figure}
\begin{figure}
\begin{indented}
\item[]\includegraphics[width=0.8\textwidth]{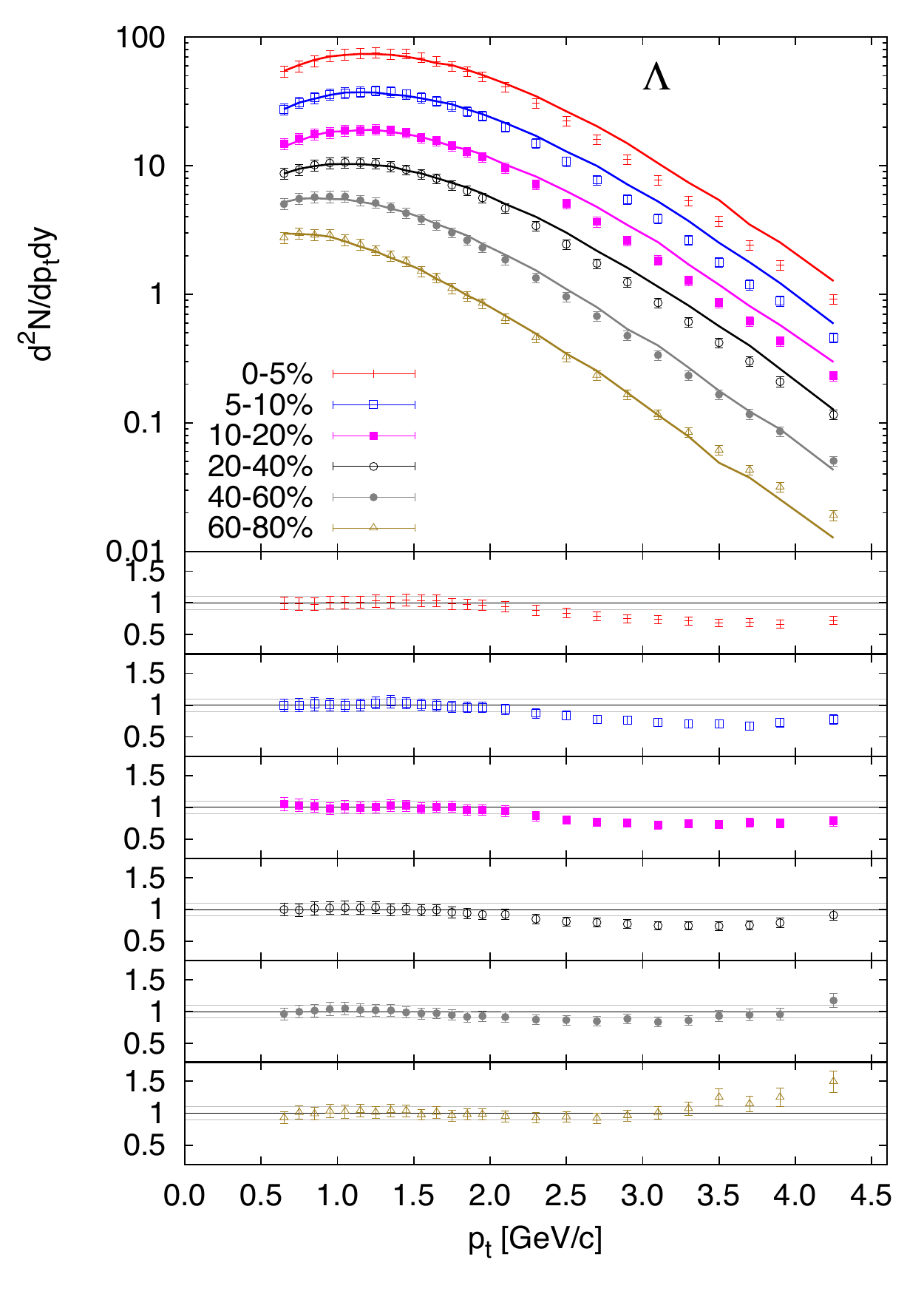}
\end{indented}
\caption{\label{f:Lfit}%
Same as Fig.~\ref{f:pifit}, but for $\Lambda$. 
}
\end{figure}
We illustrate the quality of the fits in Figs.~\ref{f:pifit}--\ref{f:Lfit}.
For brevity, for the most abundant spectra we only show the comparison with 
positively charged species and note that the comparison looks the same 
for negative species. For pions (Fig.~\ref{f:pifit}) we clearly see the enhancement 
at low $p_t$. Recall that our fitting procedure does not include any chemical 
potential  but we account for resonance production. 
For pions at $p_t$ below 400~MeV/$c$, resonance contribution grows as 
$p_t$ is lowered. This can be observed in Fig.~\ref{f:dirres}. Still, although
qualitatively this behaviour might agree with what is seen in data, quantitatively 
it is not sufficient to explain them.  
Resonances tend 
to populate low-$p_t$ region but this is not enough to fit the data. We speculate 
that the solution might be in introducing non-equilibrium chemical potential 
for pions. This would naturally occur if the hadron gas chemically freezes out 
at a temperature around 150--160~MeV and then cools down while keeping the 
effective ratios of individual species constant. We estimated that the pion chemical 
potential at kinetic freeze-out temperature might reach values around 100~MeV. 
This is not enough for Bose-Einstein condensation but modifies the spectrum considerably. 

In the higher-$p_t$ region the pion spectrum is well reproduced up to about 2~GeV. This seems 
reasonable, as for higher $p_t$ we may see signs of hard production. 

Charged kaons (Fig.~\ref{f:Kfit}) are well reproduced in a similar $p_t$ interval without 
the need to leave out any bins at low $p_t$. We also observe good fits to 
(anti)proton spectrum (Fig.~\ref{f:pfit}) stretching out even to about 4~GeV.
In general, the agreement becomes slightly worse when going away from central 
collisions. For the most peripheral class charged kaons depart from the theoretical 
curve above 2~GeV/$c$ and protons above 3.5~GeV/$c$.

In the fit we also included strange $V^0$'s which were measured up to $p_t$ 
of 4.5~GeV/$c$ \cite{ALICE_Lambda}. The kaons show (Fig.~\ref{f:K0fit}) 
similar behaviour as their
charged isospin partners.
When comparing the two plots for $K^+$ (Fig.~\ref{f:Kfit}) and $K^0$
(Fig.~\ref{f:K0fit}) one should notice the different $p_t$ intervals in which these 
spectra are measured. 
Below the  $p_t$ cut of 2~GeV also the 
$\Lambda$'s are fitted well (Fig.~\ref{f:Lfit}). They slightly depart from this agreement above 2~GeV, 
but in a different way than other species: $\Lambda$'s are steeper at high $p_t$
than the thermal fits.  Note that the values of freeze-out parameters were determined
in common fits to all 8 species, including $\Lambda$'s. We might be seeing here
the beginning of the departure of strange baryon spectra from the scenario of 
common freeze-out. 
Note, however, that the departure occurs for $p_t$ above 2~GeV/$c$ where 
thermal hydrodynamically inspired model should not be given too much stress.

\begin{figure}[t]
\begin{indented}
\item[]\includegraphics[width=0.8\textwidth]{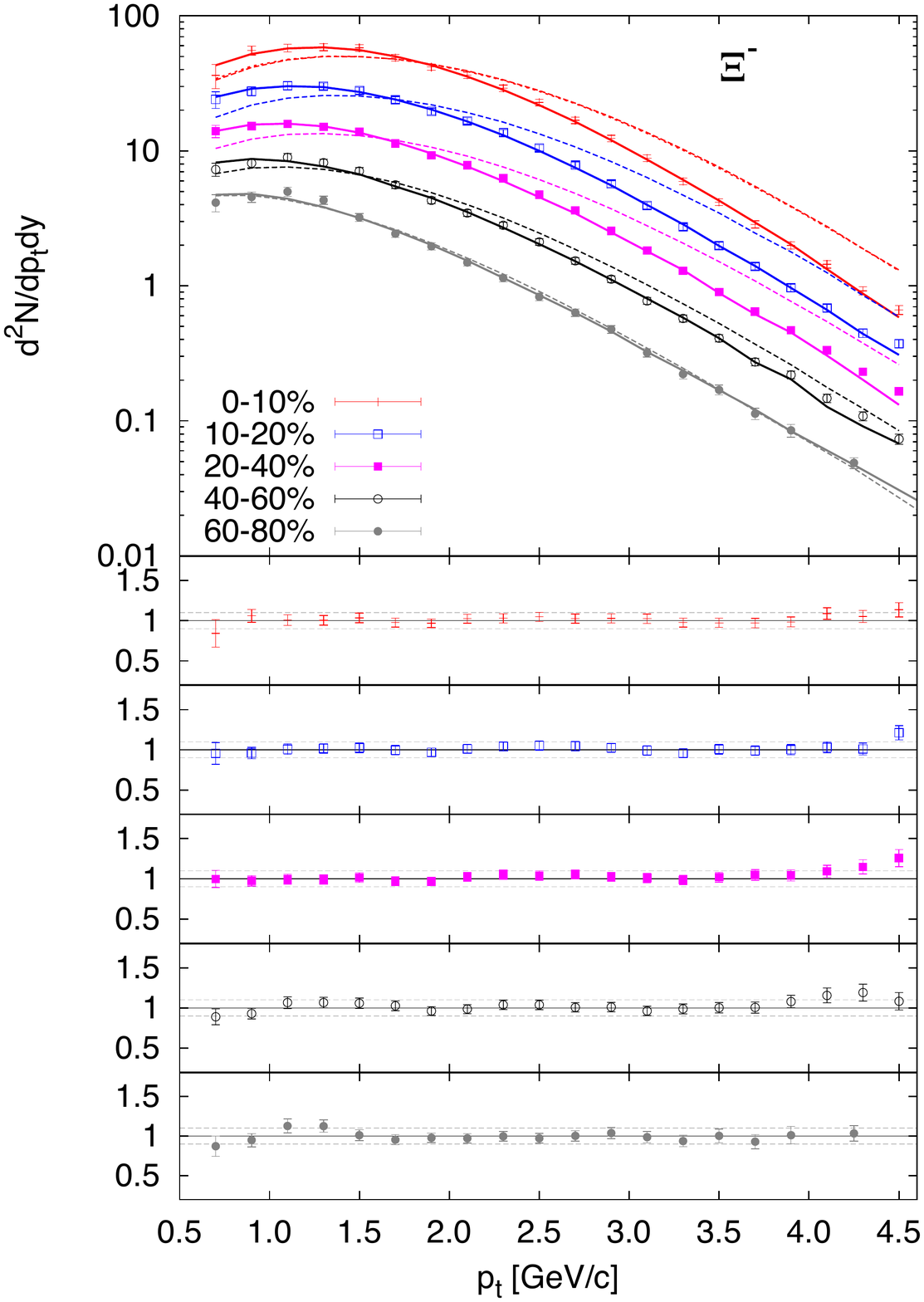}
\end{indented}
\caption{\label{f:Xi}%
Transverse momentum spectra of $\Xi^-$ for centralities indicated in 
Table~\ref{t:strangeparams}. 
To display all spectra in one figure we divide data from non-central collisions
by factors 2, 4, 8,  and 16. Dashed curves show predictions based on parameters 
from Table~\ref{t:bestfits}. For the most central bin we show two predictions 
calculated with parameters for centrality classes 0--5\% and 5--10\%; 
they differ very little. 
Lower panels show the experiment-to-Monte-Carlo ratios 
$R_i = N_i^\mathrm{exp}/N_i^\mathrm{MC}$ which demonstrate the 
agreement of measured and fitted spectra. Different panels correspond to 
different centralities. Horizontal lines indicate the interval 0.9 $\le R_i \le$ 1.1 to 
which we limit our fitting procedure.}
\end{figure}
\begin{figure}[t]
\begin{indented}
\item[]\includegraphics[width=0.8\textwidth]{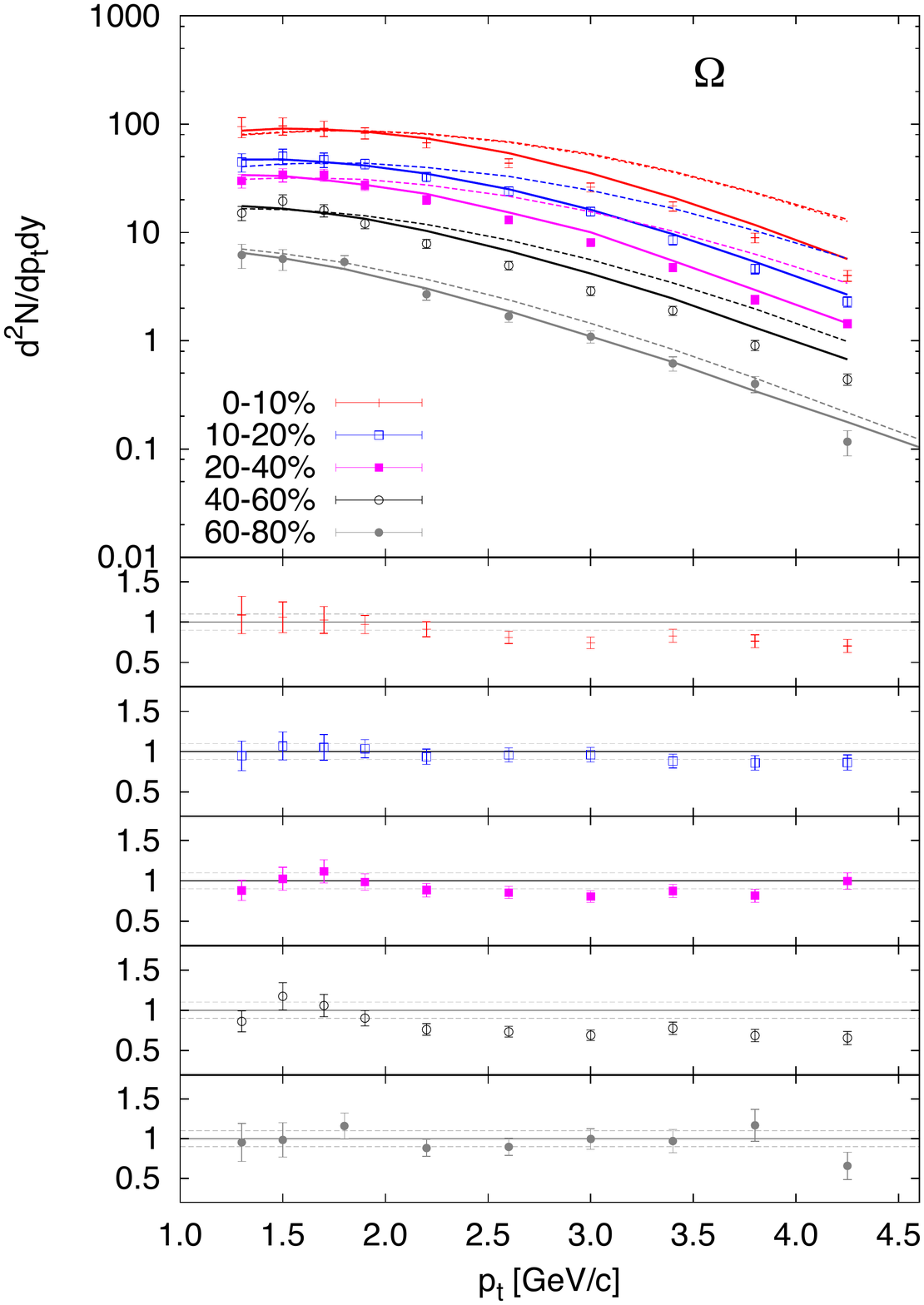}
\end{indented}\caption{\label{f:Omega}%
Same as \ref{f:Xi} but for $\Omega$.}
\end{figure}


\subsection{Spectra of $\Xi^-$, $\bar\Xi^-$, $\Omega$, and $\bar \Omega$}

The departure of data from the model fitted in previous subsection
is clearly seen when we focus on $p_t$ spectra of multiply strange baryons:
$\Xi^-$, $\bar\Xi^-$, $\Omega$, and $\bar \Omega$. We show only the spectra of baryons 
in Fig.~\ref{f:Xi} ($\Xi^-$) and Fig.~\ref{f:Omega} ($\Omega$). Results for antibaryons 
are similar. Note that these species were \emph{not} included in the previous 
fitting procedure. In order to check whether they fit into the same description of 
the fireball we have simulated their spectra in DRAGON with the sets of 
parameters that come from the fits to $\pi$, $K$, $p$, $\bar p$, $\Lambda$
(Table~\ref{t:bestfits}). These predictions are plotted in Figs.~\ref{f:Xi} and 
\ref{f:Omega} by dashed curves. We notice that in general the data are steeper 
than the prediction. The data would thus indicate lower transverse expansion 
velocities and earlier freeze-out of multiply strange baryons \cite{vanHecke:1998yu}. 
Note, however, that for the two most 
peripheral classes the agreement of data and prediction becomes better. 

In order to quantify this statement we have performed common fits to the four multistrange 
baryon species: $\Xi^-$, $\bar\Xi^-$, $\Omega$, and $\bar \Omega$ and extracted 
their values of $T$, $\eta_f$, $n$ and corresponding $\langle v_t \rangle $. They are listed in 
Table~\ref{t:strangeparams} and compared with the other fit results 
in Fig.~\ref{f:bestfit}.
\begin{table}
\caption{The best parameter values resulting from fits to transverse momentum 
spectra of $\Xi$ and $\Omega$. 
\label{t:strangeparams}}
\lineup
\begin{indented}
\item[]\begin{tabular}{lp{2.7em}lllll}
\br
centrality & $T$ (MeV)  & $\eta_f$ & $n$ & $\langle v_t \rangle$ & $\chi^2/N_\mathrm{dof}$ & $N_\mathrm{dof}$ \\
\mr
\00--10\%   & 126  & 0.82  & 0.71  & 0.605 & 0.375 & 41\\
10--20\% &  138 &  0.81 &  0.85 &  0.568 & 0.325 & 41 \\
20--40\% &  158 & 0.78 & 0.91 & 0.536 & 0.417 & 37 \\
40--60\% & 174 & 0.76 & 1.11 & 0.489 & 0.497 & 33\\
60--80\% & 250 & 0.64 & 1.35 & 0.382 & 0.795 & 39 \\
\br
\end{tabular}
\end{indented}
\end{table}
This clearly shows the general effect that multiply strange baryons seem to decouple at higher 
temperature and weaker transverse expansion. The fitted spectra are 
compared to data in Figs.~\ref{f:Xi} and \ref{f:Omega} by solid lines. Due to 
higher statistics, fit results seem to be driven by the $\Xi$'s rather than 
by $\Omega$'s. Although it is hard to support it statistically, it seems that the 
$\Omega$ spectrum is even steeper than the fit results. This would indicate even 
lower expansion velocity at the moment of $\Omega$ decoupling. 

Looking at the most peripheral centrality class we notice that the measured data 
seem to be well reproduced by both sets of fit results. Note that the freeze-out 
temperature in peripheral collisions is higher than in the central ones. The spectra in 
those collisions are thus less sensitive to the mass of particles. What we see
here is an illustration of the ambiguity
in determination of $T$ and $\langle v_t \rangle$. Lower temperature can be 
compensated by stronger flow and vice-versa. When spectra depend on 
the mass, this ambiguity is usually resolved by 
different shapes of $p_t$ spectra of different species. Figure~\ref{f:bestfit} indicates
that statistically one can still resolve the fit to multistrange baryons as the uniquely 
correct result.


\subsection{Spectra of $K^*$ and $\phi$}

\begin{figure}[t]
\begin{indented}
\item[]\includegraphics[width=0.8\textwidth]{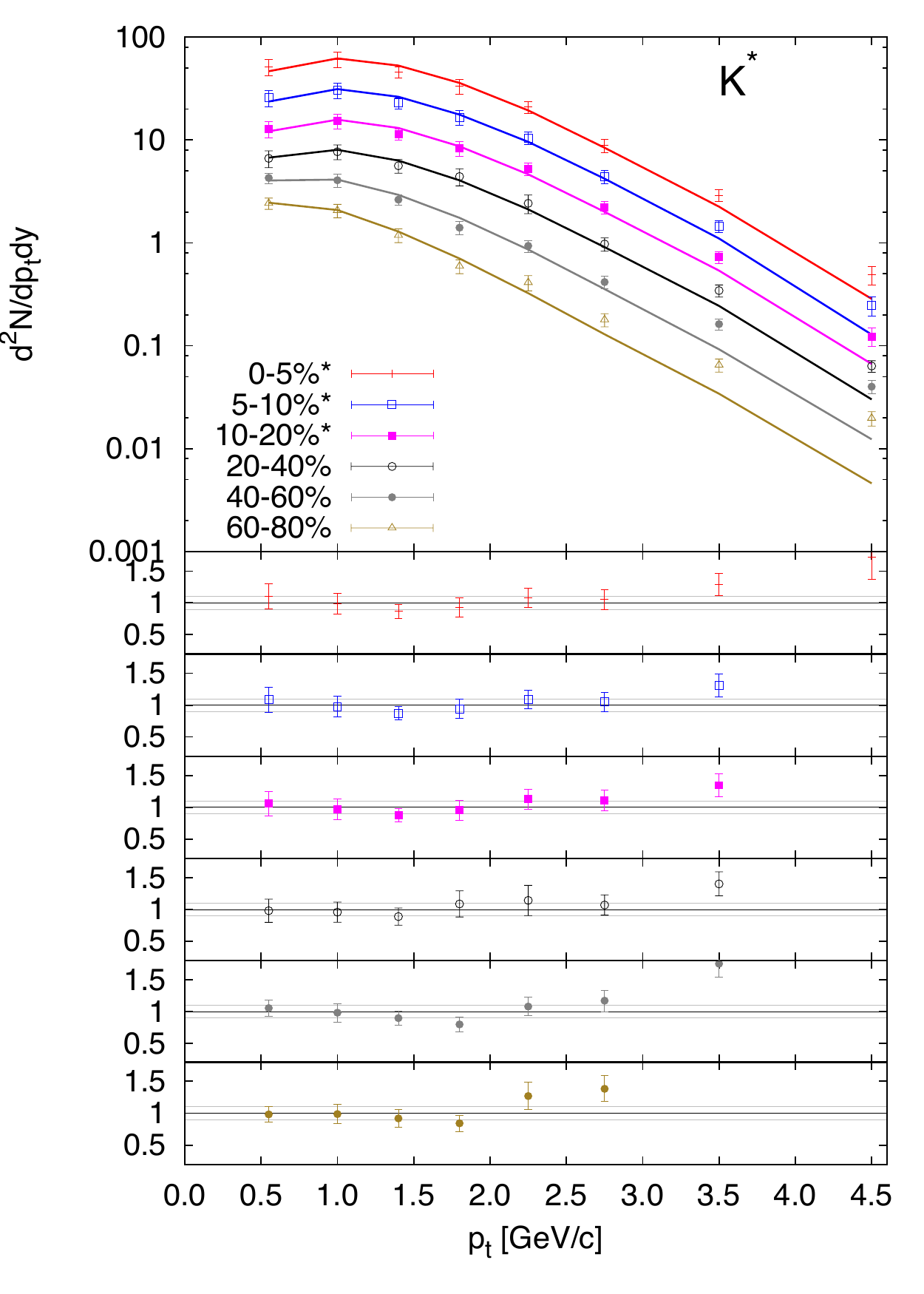}
\end{indented}
\caption{\label{f:Kstarfit}%
Comparison of measured $K^*$ spectra \cite{Abelev:2014uua} with the predictions based on 
fitted parameters from Table \ref{t:bestfits}. Stars in the legend indicate that the 
centrality 
class of data 0--20\% is compared to predictions for 0--5\%, 5--10\%, and 
10--20\%. 
To display all spectra in one figure we divide data from non-central collisions
by factors 2, 4, 8,  16, and 32.
Lower panels show the ratios $R_i = N_i^\mathrm{exp}/N_i^\mathrm{MC}$.
}
\end{figure}

\begin{figure}
\begin{indented}
\item[]\includegraphics[width=0.8\textwidth]{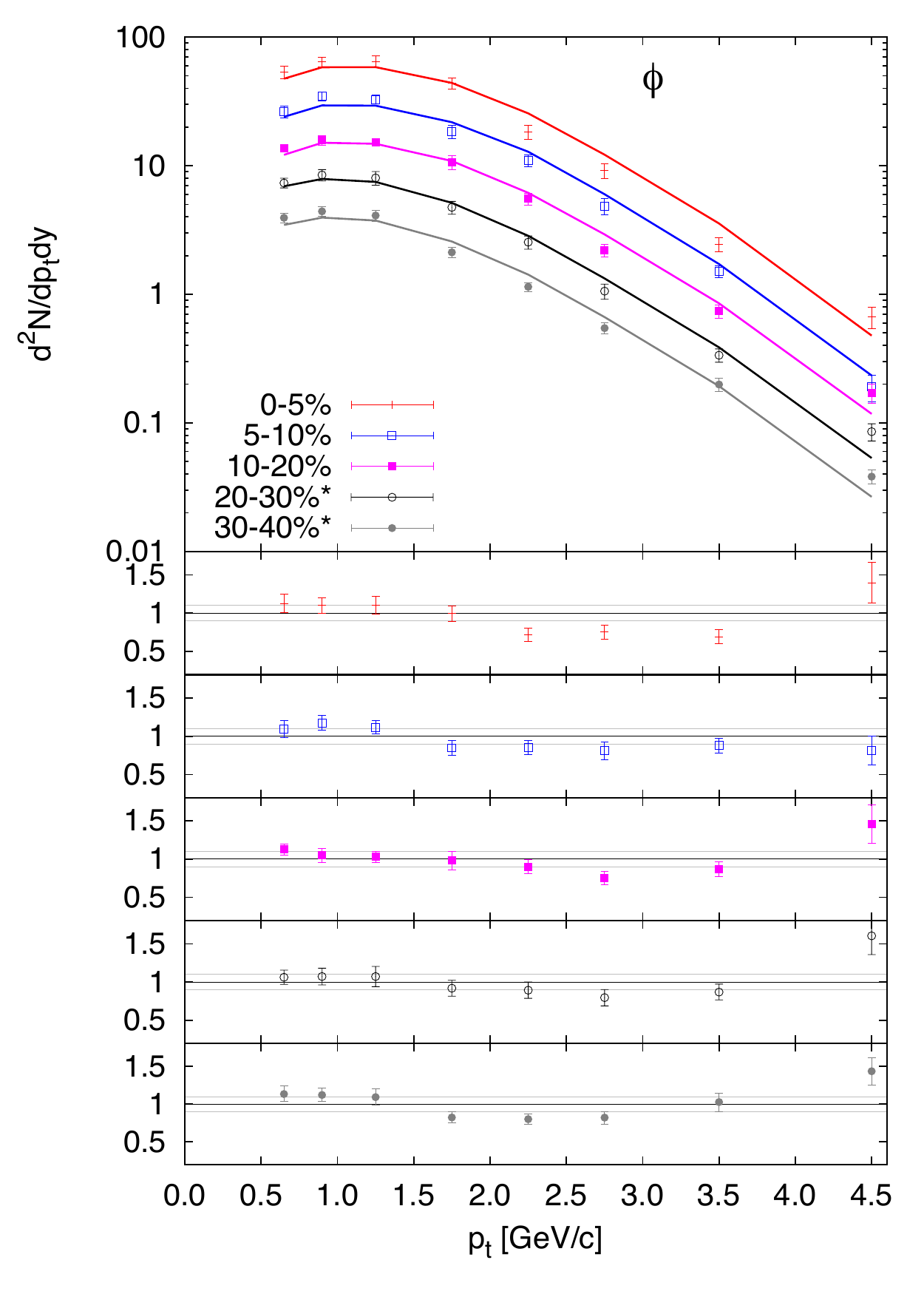}
\end{indented}\caption{\label{f:phi}%
Comparison of measured $\phi$ spectra \cite{Abelev:2014uua} with the predictions based on 
fitted parameters from Table \ref{t:bestfits}. Stars in the legend indicate that 
the centrality classes 20--30\% and 30--40\% 
of experimental data are compared to predictions 
based on the fit for 20--40\% centrality in pions, kaons and (anti)protons.
}
\end{figure}

ALICE collaboration has also published transverse momentum spectra of resonances
$K^*$ and $\phi$ \cite{Abelev:2014uua}. Studying resonances may reveal information
about the dynamics of the fireball, particularly its evolution between chemical and thermal 
freeze-out. In order to simulate these resonances we have 
forbidden them to decay in our Monte Carlo generator since by default their decay is 
turned on. We simulated their spectra with the parameters from Table~\ref{t:bestfits}
and compare them with data in Figs.~\ref{f:Kstarfit} and \ref{f:phi}. Hence, they have 
\emph{not} been fitted in our treatment. We observe reasonably good agreement 
of theory with data in centrality classes 0--20\% and  20--40\% and $p_t$ below
2~GeV/$c$ for the $K^*$. For higher $p_t$ the data tend to be flatter than the 
theory. 

The $\phi$ meson with its mass 1.02~GeV/$c^2$ is an important probe of 
fireball dynamics. If it is produced from the thermalized fireball at the common 
freeze-out, then its spectrum must be close to that of protons. In Fig.~\ref{f:phi}
we compare the data with predictions based on parameters from Table~\ref{t:bestfits}. 
Data seem to be slightly steeper than the prediction. This is best seen in the ratios
$R_i$ plotted in the lower panels of the Figure. Such a disagreement might indicate 
slightly earlier freeze-out from the fireball. Note, however, that the deviation of 
the prediction from the data is much smaller than in the case of multistrange baryons and 
no indisputable sign of earlier decoupling can be claimed.


\section{Conclusions}
\label{s:conc}

The blast-wave model with hadron production from resonance decays included 
can well reproduce single-particle 
spectra measured by ALICE collaboration in Pb+Pb collisions at 
$\sqrt{s_{NN}} = 2.76$~TeV. 
Resonance contribution leads to shifts in the values of freeze-out parameters. For central collisions our freeze-out temperatures
tend to be lower than those extracted by ALICE \cite{ALICE_piKp}. On the other hand, the temperatures shift 
in the opposite direction for peripheral collisions. Shifts in $\langle v_t \rangle$ have smaller relative size.

Multistrange baryons show freeze-out at higher temperature and weaker transverse flow. 
This is consistent with earlier decoupling from the fireball. Note that such a behaviour 
is known from SPS \cite{vanHecke:1998yu} and RHIC \cite{Adams:2003fy} data, as well. 

In fits to data from most peripheral collisions we obtain temperatures for kinetic freeze-out 
that are higher than what is usually assumed for the chemical freeze-out. On one side 
we can take this result as a parametrisation and the values of parameters are those which 
fit the data best. However, as soon as we want to interpret these values in the framework of 
some scenario for the fireball evolution, we may be witnessing limitations of the used model. 

In principle, our assumption of chemical equilibrium may seem inconsistent with individual normalisation of each spectrum as we performed it. If different spectra are multiplied with 
different factors this can be seen as effectively changing relative numbers of final state 
species. Caution would be particularly 
in place if we tried to predict the absolute yields of final hadron species. 
Here, however, we wanted to enhance the sensitivity to $T$ and $\langle v_t\rangle$ and 
thus we focussed on the shapes of the spectra which remain intact by the individual normalizations. Indeed, although rescaling one spectrum might have an influence on 
another one through changing yields of resonances which also contribute to a different  
final state species (like e.g.\ $\Delta$ contributing to pions and nucleons), this does not 
tie different spectra too strongly together. The reason is that 
 the main contributing resonance decays are different for 
different hadron species: for pions  the most important are $\rho$, $\omega$, 
$\eta$ \cite{sqmPROC}; 
for protons the $\Delta$ resonance; for $\Lambda$ the $\Sigma^0$ and $\Sigma^*$ hyperons; 
for kaons the $K^*$. Thus rescaling the yield for one species does not have much influence 
on another species. 

A more complete analysis would include also the model interpretation of the 
normalisations which are given by the volume. Such an analysis, however, must 
include femtoscopy and is planned for the future. 

Similar fits were performed recently with the Cracow single freeze-out model 
\cite{Begun:2013nga,Begun:2014rsa}. The main feature of that model is common chemical and 
thermal freeze-out with non-equilibrium fugacities introduced for strange and 
also non-strange species. Due to this, the freeze-out temperature fitted there is
generally higher than our results. Note however, that  their freeze-out 
hypersurface has a different shape. Naturally, this opens the question if one can 
distinguish between the two models. Perhaps femtoscopy can provide the tool for
this task.

It is an intriguing speculation if the two approaches may be equivalent. Here we 
recall the scenario that matter hadronizes and cools down 
while keeping partial chemical equilibrium. 
This leads to the appearance of non-equilibrium chemical 
potentials for all species, i.e.\ chemical potentials that are not given by chemical 
potentials of conserved quantum numbers \cite{Bebie:1991ij}. 
In the single freeze-out model, on the other hand,  the system 
is born into the chemical non-equilibrium. We wonder how close to this non-equilibrium 
state the fireball comes in the cooling fireball scenario.

A comprehensive picture of the gradual fireball cooling should also include earlier 
decoupling of multistrange baryons.

Finally, let us come back to the issue that our 
results have been obtained in a scenario which 
assumes different chemical and thermal freeze-out and the difference between 
their temperatures is rather high. If one would  revisit the question of the 
evolution between them,  the inclusion of chemical 
potentials may be important. 
They may have influence on the shape of $p_t$ spectra especially for 
pions and we seem to observe the corresponding features in the data. This task will be 
addressed in a subsequent paper.   



\ack
We thank Evgeni Kolomeitsev, Karel \v{S}afa\v{r}\'ik, and J\"urgen Schukraft for valuable discussions.
We gratefully acknowledge financial support by grants 
APVV-0050-11, VEGA 1/0457/12 (Slovakia) and 
M\v{S}MT grant  LG13031 (Czech Republic). 
Computing was performed in the High Performance Computing Center of the Matej Bel University in Bansk\'a Bystrica using the HPC infrastructure acquired in project ITMS 26230120002 and 26210120002 (Slovak infrastructure for high-performance computing) supported by the Research \& Development Operational Programme funded by the ERDF.


\end{document}